\documentstyle[epsf,12pt]{article}

\begin{document}
\baselineskip 1.5em		
\parindent 2em


\title{Recent K2K Results
\footnote{\uppercase{K}\uppercase{E}\uppercase{K} \uppercase{P}reprint 2004-22}
\footnote{\uppercase{T}alk presented at 
the 5th \uppercase{W}orkshop on \uppercase{N}eutrino 
\uppercase{O}scillations and their \uppercase{O}rigin 
(\uppercase{N}\uppercase{O}\uppercase{O}\uppercase{N}2004), 
\uppercase{T}okyo, \uppercase{J}apan, \uppercase{F}ebruary 11-15, 2004}}

\author{Takanobu ISHII\\
\footnotesize{(for the K2K Collaboration)}\\
\footnotesize{KEK/IPNS, 1-1 Oho, Tsukuba, Ibaraki, JAPAN}\\
\footnotesize{E-mail: takanobu.ishii@kek.jp}
\footnote{\uppercase{P}artially
supported by the \uppercase{G}rants for \uppercase{S}cientific 
\uppercase{R}esearch of \uppercase{M}onkasho.} \\
}
\date{}

\maketitle

{\footnotesize
The disappearance of $\nu_{\mu}$ was studied using the K2K-I dataset, 
which was taken before July, 2001. 
We observed indications of neutrino oscillation. 
The resultant oscillation-parameter region was consistent with 
the atmospheric neutrino result. 
The appearance of $\nu_{e}$ was searched for in the same dataset.  
No excess was found over the expected background. }

\section{Introduction}
K2K\cite{K2K} (KEK to Kamioka long-baseline neutrino experiment) 
is the first accelerator-based long-baseline neutrino experiment 
to investigate the neutrino oscillation observed in atmospheric 
neutrinos.\cite{SKevi} 
K2K intends to confirm the $\nu_{\mu}$ disappearance by observing a 
reduction of $\nu_{\mu}$ events and a distortion of the $E_{\nu}$ 
spectrum, and to search for the appearance of $\nu_{e}$.

\section{Beam and Detectors}
For neutrino-beam production, 12-GeV protons from the KEK 
proton synchrotron (PS) are extracted 
in a $1.1\mu$s spill every 2.2seconds. 
The proton beam is bent to the Kamioka direction, and hits an 
aluminum target embedded in the first horn. 
A pair of horn magnets focuses positively charged particles, mainly 
positive pions produced in the proton-aluminum interaction.  
A pion monitor\cite{PIMON} is occasionally put downstream of the 2nd horn 
in order to measure the direction and momentum of the pions. 
The positive pions decay into $\mu^{+}$ and $\nu_{\mu}$ 
during their flight in a 200-m 
decay pipe.  The $\nu_{\mu}$ is used for the experiment.  
The beam is monitored by a muon monitor pulse-by-pulse 
by measuring the muons from $\pi$ decay.  
The neutrino beam, itself, is monitored by a near detector system 
located 300 m from the target.  

The near detector system consists of a 1-kiloton water cherenkov detector 
(1KT), a scintillation fiber detector (SCIFI),\cite{SCIFI} 
a muon range detector 
(MRD)\cite{MRD} and originally a lead-glass detector, which was replaced 
with a new scintillator-bar detector (SCIBAR) last summer. 

K2K uses Super-Kamiokande (SK) as the far detector, which is 
located at 250 km away.  

As for the data-taking period, before the SK restoration we call it K2K-I 
and after SK restoration it is called K2K-II.  
K2K-II started in December, 2002. 
SCIBAR was installed after half-year running of K2K-II.

\section{Data accumulation and data quality}
Even during the K2K-II period, we have been accumulating data smoothly. 
The delivered protons on target (POT) reached $9.5\times10^{19}$ 
at the end of the year 2003. 

%
\begin{figure}[ht]
\centerline{\epsfxsize=4.1in\epsfbox{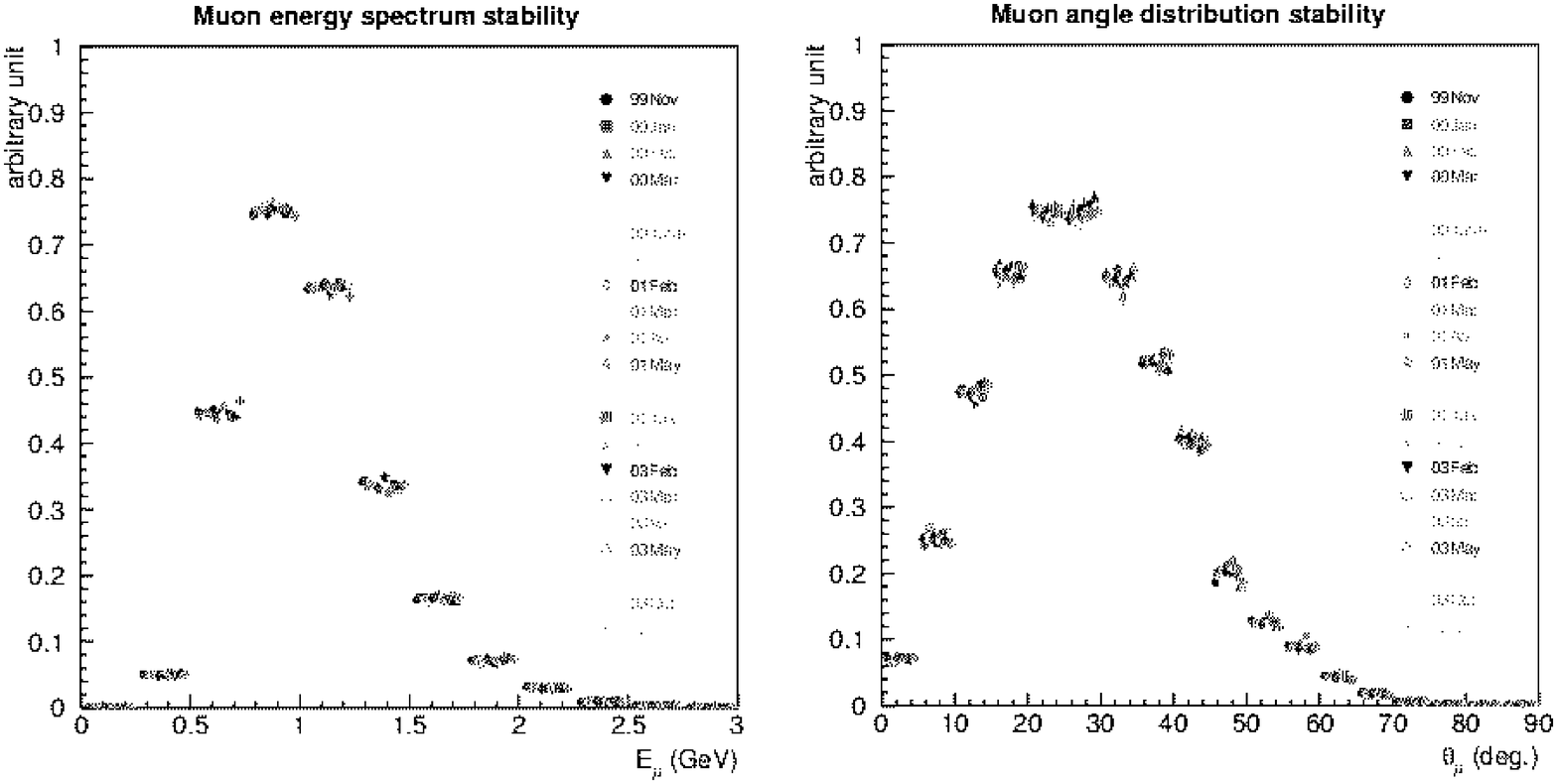}}   
\caption{(left): Stability of the muon-energy distribution plotted every month.
 (right): Stability of the muon-angle distribution plotted every month.
 \label{emu-amu}}
\end{figure}

\begin{figure}[ht]
\centerline{\epsfxsize=3.0in\epsfbox{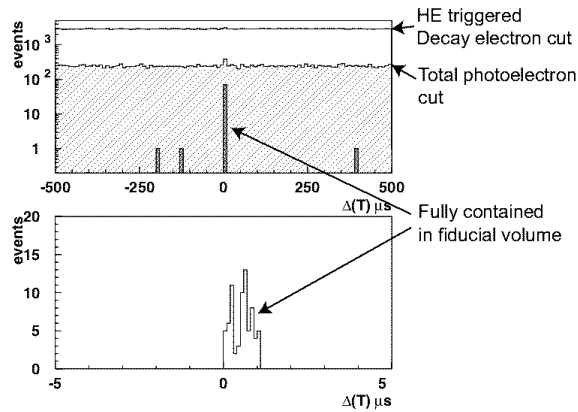}}   
\caption{(top): $\Delta T$ distribution after each cut. 
 (bottom): Expanded $\Delta T$ distribution for the final sample.
 \label{deltaT}}
\end{figure}

\begin{figure}[ht]
\centerline{\epsfxsize=4.1in\epsfbox{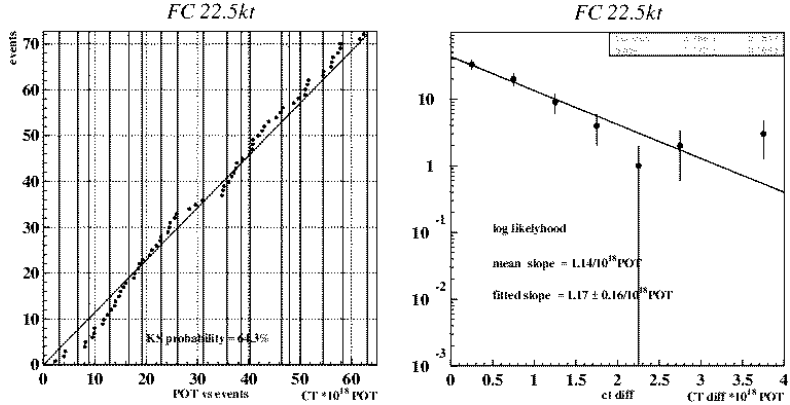}}   
\caption{(left): SK event number vs. POT.
  (right): Interval distribution between two consecutive events. 
        The line shows the result of the exponential fit.
 \label{ctev-gap}}
\end{figure}

The neutrino direction is stable within 1 mrad for 
the entire experimental period, which has been monitored by MRD.  
According a beam Monte Carlo simulation, the neutrino flux at SK does not 
change by more than 1\%, even in a beam direction shifted by 3 mrad. 
The neutrino spectrum was measured by the pion monitor at 
the begginning.  The spectrum stability has been confirmed by measurements 
of the energy and angle of muons produced by the charged-current 
interactions in the MRD.  Fig.~\ref{emu-amu} shows the stability 
of the muon energy spectrum and the muon angle plotted monthly. 

SK events generated by the K2K neutrinos are selected in the same 
way as in the atmospheric neutrino analysis.  
Fig.~\ref{deltaT} shows the remaining events after each cut as a function 
of $\Delta T$.  $\Delta T$ is the difference of the GPS time stamps 
between the SK event and the beam spill time.  
In this figure, the data sample of June'99 to April'03 is plotted, 
namely K2K-I plus part of K2K-II.  
We observe 72 events in the $1.5\mu$s time window.  
The expected atmospheric neutrino background in this time window 
is about $2\times10^{-3}$ events.  
The selected 72 events are plotted as a function of POT in 
Fig.~\ref{ctev-gap} (left).  
During the K2K-II period, the event rate seems to be the same as that 
of K2K-I.  Fig.~\ref{ctev-gap} (right) shows the event-gap 
distribution of each 
consecutive event.  It fits to an exponential shape quite well, 
as expected.  This means that the fluctuation of the event rate 
is statistical.  

\section{Oscillation analysis}
For an oscillation analysis of $\nu_{\mu}$ disappearance,\cite{K2Kind} 
we use both the number of events and the spectrum shape distortion 
in the K2K-I dataset.  
For the spectrum analysis, first we deduced the spectrum 
at the near site using the 1KT and SCIFI data.  
The pion monitor data was used to constrain the fit and 
also to extrapolate the near spectrum to the far spectrum.  

\begin{figure}[ht]
\centerline{\epsfxsize=3.5in\epsfbox{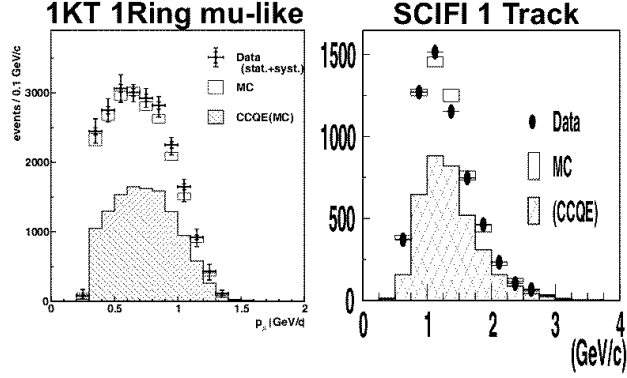}}   
\caption{(left): Momentum distribution of the 1-ring $\mu$-like events in 1KT.
  (right): Momentum distribution of the 1-track events in SCIFI which has 
  a corresponding track in MRD.
  The crosses are data and the boxes are MC with the best-fit parameters. 
 \label{ktsf1pmu}}
\end{figure}
\begin{figure}[ht]
\centerline{\epsfxsize=4.1in\epsfbox{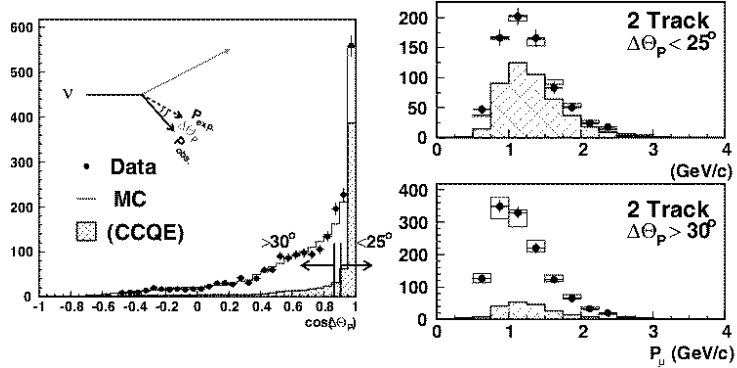}}   
\caption{(left): Cosine of $\Delta\theta_{p}$.
 (top right): SCIFI $p_{\mu}$ distribution for the QE enriched sample.
 (bottom right): SCIFI $p_{\mu}$ distribution for the non-QE enriched sample.
  The crosses are data and the boxes are MC with the best-fit parameters. 
  The hatched histogram shows the QE events estimated by MC. 
 \label{sf2track}}
\end{figure}

For 1KT, single-ring $\mu$-like events are used to enrich charged current 
quasi-elastic (QE) events.  
Since SCIFI has a good tracking ability, we use both 1-track 
and 2-track events.  
The momentum distributions of the 1KT singe-ring $\mu$-like events 
and the SCIFI 1-track events are plotted in Fig.~\ref{ktsf1pmu}. 
For the SCIFI 2-track events, as is shown in Fig.~\ref{sf2track}, 
we can enrich the QE or non-QE by looking 
at the $\Delta \theta_{p}$ distribution, where $\Delta\theta_{p}$ 
is the difference 
of the 2nd-track angle measured and the angle calculated from 
the 1st track assuming QE kinematics.  
The events with $\Delta\theta_{p}<25\deg$ are used as the QE 
enriched sample and those with 
$\Delta\theta_{p}>30\deg$ 
are used as the non-QE enriched sample.  
These 4 event categories are fit simultaneously.  
The fitting parameters are 8 $E_{\nu}$ bins and non-QE/QE ratio.  
After the fit, the agreement between data and MC looks good for 
all the categories.  

\begin{figure}[ht]
\centerline{\epsfxsize=2.5in\epsfbox{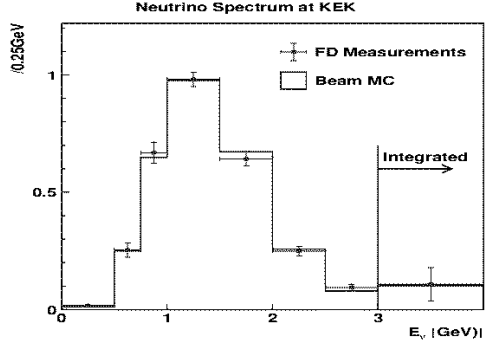}}   
\caption{Fitted $E_{\nu}$ spectrum at the near site.
 \label{fdspec}}
\end{figure}

The result for the $E_{\nu}$ spectrum at the near site is given in 
Fig.~\ref{fdspec} compared with the beam MC.  

\begin{figure}[ht]
\centerline{\epsfxsize=4.1in\epsfbox{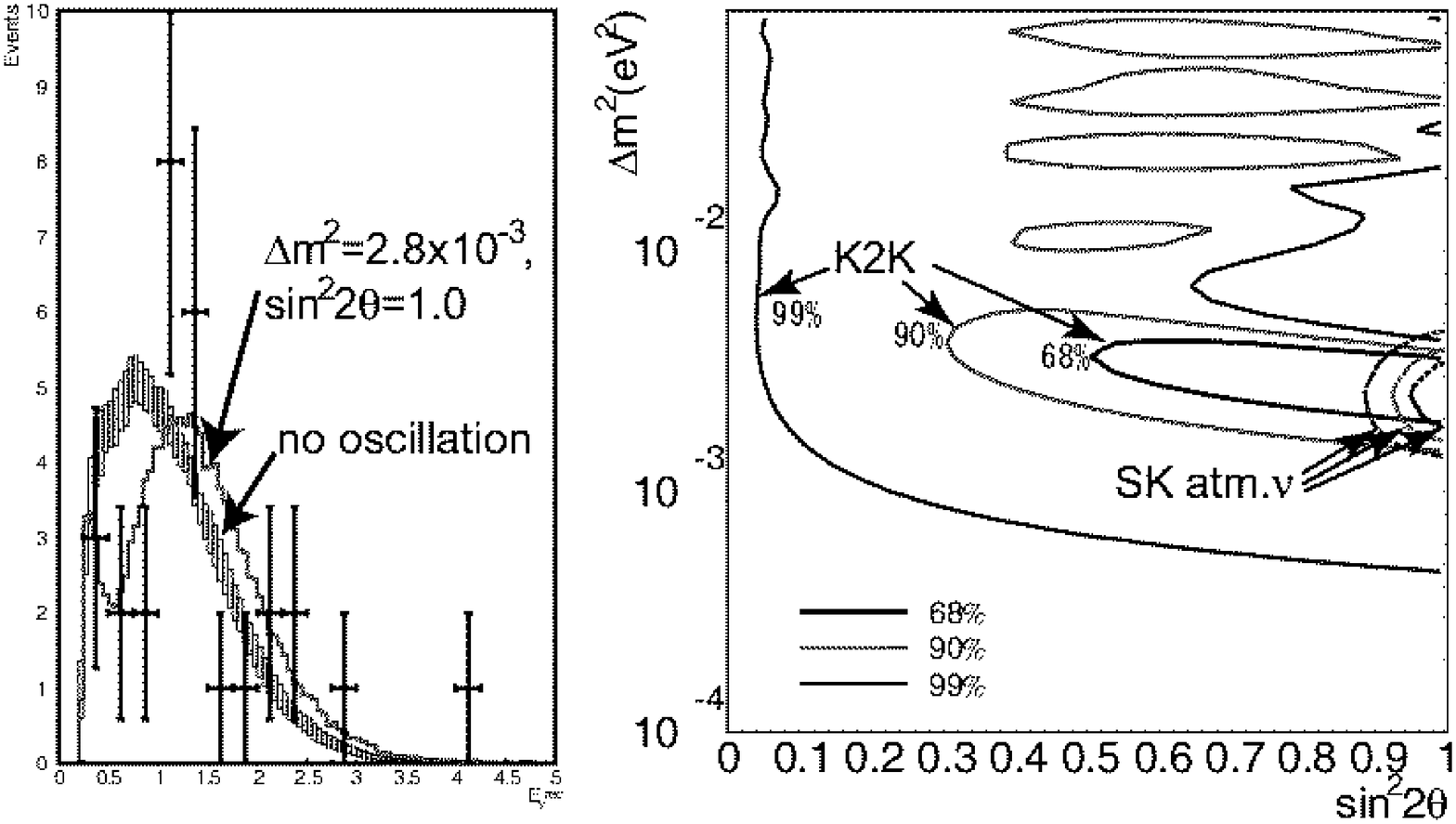}}   
\caption{(left): Reconstructed neutrino energy distribution at SK.
The boxes show the expectation from the no-oscillation case and the solid 
histogram shows the best-fit result.  
Normalization is by area.  
 (right): Allowed oscillation parameter regions.
 \label{numuresult}}
\end{figure}

We observed 56 events at SK in the K2K-I 
dataset, while the expectation from the near site measurement 
is $80.1^{+6.2}_{-5.4}$.  
Figure~\ref{numuresult} (left) is the reconstructed $E_{\nu}$ distribution 
of the single ring $\mu$-like events at SK 
compared with the expectation from the near site measurement. 
From an analysis using both the number of events and 
the spectrum shape, the null oscillation probability is less than 1\%.  
Fig.~\ref{numuresult} (right) shows the resultant allowed parameter region.

\section{Search for $\nu_{e}$ appearance}
\begin{table}[ph]
\centering
\caption{Reduction summary of the $\nu_{e}$ appearance search. 
$\nu_{\mu}$ MC shows the background from misidentified $\nu_{\mu}$ 
interactions, assuming that no oscillation exists.  The last 
column is the expected signal, assuming $sin^{2}2\theta_{\mu e}=1$, 
and $\Delta m^{2} = 2.8\times 10^{-3}eV^{2}$.}
{\footnotesize
\begin{tabular}{@{}ccccc@{}}
\hline
{} &{} &{} &{} &{}\\[-1.5ex]
{} & Data & $\nu_{\mu}$ MC & Beam $\nu_{e}$ MC & Signal $\nu_{e}$ MC(CC)\\[1ex]
\hline
{} &{} &{} &{} &{}\\[-1.5ex]
FCFV              &56 &80  &0.82 &28\\[1ex]
Single ring       &32 &50  &0.48 &20\\[1ex]
PID(e-like)       &1  &2.9 &0.42 &18\\[1ex]
$E_{vis}\>100MeV$ &1  &2.6 &0.41 &18\\[1ex] 
w/o decay-e       &1  &2.0 &0.35 &16\\[1ex]
\hline
\end{tabular}\label{tab-nuereduc} }
\vspace*{-13pt}
\end{table}

\begin{figure}[ht]
\centerline{\epsfxsize=4.1in\epsfbox{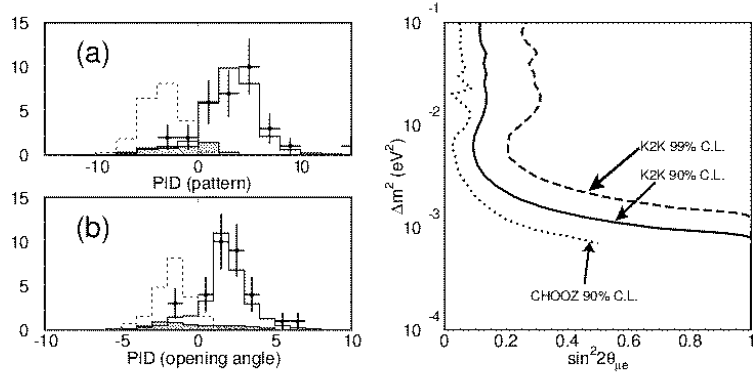}}   
\caption{(left): PID likelihood distributions. 
The top figure is based on the ring pattern.  
and the bottom figure is based on the opening angle.  
The solid histogram is $\nu_{\mu}$ MC and the shaded area is 
the neutral-current 
component of the $\nu_{\mu}$ MC. 
The dotted histogram is the signal $\nu_{e}$ MC, assuming 
$sin^{2}2\theta_{\mu e}=1$ 
and $\Delta m^{2} = 2.8\times 10^{-3}eV^{2}$.  
 (right): Excluded $\nu_{\mu} \rightarrow\nu_{e}$ oscillation parameter 
 regions.
 \label{nueresult}}
\end{figure}

Appearance of $\nu_{e}$ is searched for\cite{K2Knue} in the K2K-I dataset. 
We selected fully contained single ring, e-like events using 
the ring pattern and the opening angle.  Then, the visible-energy cut 
and the decay-electron cut were placed.  
Fig.~\ref{nueresult} (left) shows the PID likelihood distributions. 

Table~\ref{tab-nuereduc}  gives a reduction summary.  
We have obtained 56 fully contained events at SK.  After the selections, 
only 1 event remains.  
The main background comes from the $\nu_{\mu}$ interactions, of which 
the neutral-current $\pi^{0}$ production is dominant.  
We have encountered about 1\% $\nu_{e}$ contamination in 
the $\nu_{\mu}$ beam, 
which also becomes the background.  
The observed 1 event is consistent with the expected 
background.  
The result of $\nu_{e}$ appearance is drawn in Fig.~\ref{nueresult} (right).  
The excluded region is given by the K2K.  The dotted line is 
the CHOOZ excluded region, 
which measured the $\bar{\nu_{e}}$ disappearance.  
The CHOOZ result was converted into $sin^{2}2\theta_{\mu e}$, 
assuming full mixing in the 2-3 sector.

\section{New near detector SCIBAR}
\begin{figure}[ht]
\centerline{\epsfxsize=4.1in\epsfbox{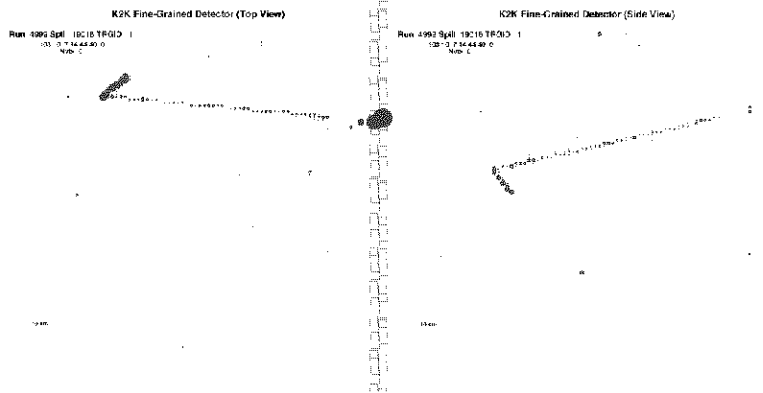}}   
\caption{Example of a SCIBAR event.
  The size of each point shows the pulse height of each scintillator bar.  
  The left figure is a top view and the right one is a side view. 
 \label{sbevent}}
\end{figure}

In order to explore the low-energy region, a new near detector, SCIBAR, 
was installed during the summer of 2003.  SCIBAR consists of 15,000 
scintillator bars read out by the WLS fibers.  
Fig.~\ref{sbevent} shows an example of a SCIBAR event.

\section{Summary}
The K2K-I data set has been analyzed.  
The reduction of the $\nu_{\mu}$ flux together with a distortion of the energy 
spectrum was observed.  
The probability that the measurements at SK can be explained by statistical 
fluctuation is less than 1\%.  
The fitted oscillation parameters are consistent with those suggested by 
atmospheric neutrinos.  
The appearance of $\nu_{e}$ has been searched for.  
One candidate event has been found, which is consistent with the background.  
An excluded region is set for the $\nu_{\mu} \rightarrow\nu_{e}$ oscillation.  
K2K-II data taking is going on smoothly.  
The event rate is consistent with K2K-I.  
The new near detector SCIBAR is working well.  
Low-energy neutrino data is coming.

\end{document}